\documentclass[12pt,a4paper,notitlepage,fleqn]{article}

\usepackage{setspace}
\usepackage[margin=1in]{geometry}
\usepackage[symbol]{footmisc}
\usepackage[T1]{fontenc}
\usepackage[utf8]{inputenc}
\usepackage{authblk}

\makeatletter

\newcommand{\fboxsubsec}[1]{
	\begin{flushleft}
		#1
	\end{flushleft}
	}
\renewcommand{\subsection}{\@startsection{subsection}{2}{0pt}
	{1ex}
	{0.5ex}
	{\reset@font\it\fboxsubsec}
	}
\makeatother

\title{Evaluating the Financial Market Function in Prewar Japan using a Time-Varying Parameter Model}%

\author{Kenichi Hirayama$^{a}$ \ and \ Akihiko Noda$^{b,c}$\thanks{\scriptsize Corresponding Author. E-mail: noda@sanken.keio.ac.jp, Tel. +81-3-3296-4545, Fax. +81-3-3296-4545.}

{\scriptsize ${}^{a}$ \it Tokio Marine Asset Management Co., Ltd., 18th Floor Tekko Building, 1-8-2 Marunouchi, Tokyo 100-0005, Japan} 

{\scriptsize ${}^{b}$ \it School of Commerce, Meiji University, 1-1 Kanda-Surugadai, Chiyoda-ku, Tokyo 101-8301, Japan}

{\scriptsize ${}^{c}$ \it Keio Economic Observatory, Keio University, 2-15-45 Mita, Minato-ku, Tokyo 108-8345, Japan}}

\date{This Version: \today}

\renewcommand\thefootnote{\arabic{footnote}}

\usepackage{graphicx}

\usepackage[sort]{natbib}%
\usepackage{amsmath,amssymb}%
\usepackage{ascmac}%
\usepackage{multirow}%
\usepackage{lscape}%
\usepackage{subfigmat}

\usepackage{pifont}%
\usepackage{arydshln}%
\usepackage[format=hang]{caption}
\usepackage[all]{xy}
\usepackage{url}
\bibpunct{(}{)}{;}{a}{}{,}

\def\hsymbu#1{\smash{\lower1.7ex\hbox{\huge$#1$}}}

\def\ve #1{{\mbox{\boldmath $#1$}}}

\newcommand{\citetapos}[1]{\citeauthor{#1}'s \citeyearpar{#1}}
\newcommand{\citeapos}[2]{\citeauthor{#1}'s (\citeyear{#2})}
\newcommand{\ex}{{\mathbb{E}}}

\def\ve #1{{\mbox{\boldmath $#1$}}}

\begin{document}

\begin{titlepage}

\renewcommand{\thepage}{}
\renewcommand{\thefootnote}{\fnsymbol{footnote}}

\maketitle

\vspace{-10mm}

\noindent
\hrulefill

\noindent
{\bfseries Abstract:} This paper explores when the financial market lost the price formation function in prewar Japan in the sense of \citetapos{fama1970ecm} semi-strong form market efficiency using a new dataset. We particularly focus on the relationship between the prewar Japanese financial market and several government policy interventions to explore whether the semi-strong form market efficiency evolves over time. To capture the long-run impact of government policy interventions against the markets, we measure the time-varying joint degree of market efficiency and the time-varying impulse responses based on \citeapos{ito2014ism}{ito2014ism,ito2017aae} generalized least squares-based time-varying vector autoregressive model. The empirical results reveal that (1) the joint degree of market efficiency in the prewar Japanese financial market fluctuated over time because of external events such as policy changes and wars, (2) the semi-strong form EMH is almost supported in the prewar Japanese financial market, (3) \citetapos{lo2004amh} adaptive market hypothesis is supported in the prewar Japanese financial market even if we consider that the public information affects the financial markets, and (4) the prewar Japanese financial markets lost the price formation function in 1932 and that was a turning point in the market.\\



\noindent
\hrulefill

\end{titlepage}

\bibliographystyle{asa}


\section{Introduction}\label{semi_strong_sec1}

For a long time, economic historians believed that banks consistently played a central role in the Japanese financial system since the late nineteenth century (see \citet{ishii1997irj,ishii1999tfh} for details). Recent empirical studies on the prewar Japanese financial system, however, provide us a prevailing view that there was a disconnection in the system by World War I\hspace{-.1em}I. In particular, they reveal quantitatively that the prewar Japanese financial system was market-centered (not bank-centered). For instance, \citet{okazaki1993hoc} argue that the turning point was after the 1930s when the Japanese government strengthened to control the economy. \citet{hoshi2001cfg} shade light on the corporate finance and governance of the prewar Japanese company and describe that the mid-1940s was the turning point in the Japanese financial markets. \citet{teranishi2003esj,teranishi2011fsp} shows that the collapse of the major shareholders and large landowners classes and strengthening of government interventions in the 1920s led to the extinction of the Japanese financial system. Therefore, there exists a consensus on whether the disconnection has happened but there is no consensus about when it happened. In other words, it remains an open question about when the market lost price formation function in the prewar Japanese financial markets (stock, government bond, and foreign exchange).

To investigate whether the price formation function of the financial markets works or not, we often examine \citetapos{fama1970ecm} efficient market hypothesis (EMH). There are three forms of the EMH: the weak-form, semi-strong form, and strong-form. Most of the previous studies examine the weak-form and the semi-strong form EMH in the stock markets; but almost none examine the strong-form EMH. This is because there is no arbitrage opportunity even if they can utilize the inside information. If the strong-form EMH is true, investors do not have an incentive for trading. The weak-form EMH is the most well-known hypothesis that asserts that current stock prices only reflect historical prices. As \citet{lim2011esm} point out, many recent studies on the weak-form EMH conclude that stock markets are almost efficient but market efficiency changes over time. This means that there is a strong possibility that an evolutionary alternative to the EMH, the adaptive market hypothesis (AMH) of \citet{lo2004amh}, is supported in the stock markets. The AMH reinforces the view that the market evolves over time, as does market efficiency. It implies that market efficiency can arise from time to time due to changing market conditions such as behavioral bias, structural change, and external events. 

The semi-strong form EMH asserts that security prices adjust rapidly to the release of the price around new public information. We consider that new public information includes not only corporate financial information but also information on government policy interventions. In the view of corporate financial information, several studies show that financial ratios and macro variables are useful in predicting stock returns, and they conclude that the semi-strong form EMH is not supported (e.g., \citet{fama1981srr}; \citet{campbell1987srt}; \citet{campbell1988spe,campbell1988dpr}; \citet{fama1988dye,fama1989bce}). Studies in the second category, on the other hand, shed light on the relationship between the government policy interventions and stock market efficiency (e.g., \citet{davidson1982mps}; \citet{pearce1985spe}; \citet{darrat1988fps}; \citet{hancock1989fpm}; \citet{bernanke2004wpl}; \citet{ehrmann2004tsm}; \citet{laopodis2009fps}). This is because government policy interventions might cause the divergence between fundamental value and the prevailing stock price, and decline the stock market efficiency. However, whether the stock prices reflect the information on government policy interventions and whether the semi-strong form EMH is supported or not remain controversial (see \citet{fama1991ecm}, \citet{becker1996mne}, and \citet{malkiel2003emh} for details).

Furthermore, several recent studies examine how the central bank's unconventional monetary policy affects the stock market. \citet{chulia2010aef} explore the Federal Reserve's announcement effect for the federal funds target rate changes on the stock prices, volatilities, and correlations. They find that the stock market responds differently to positive and negative target rate surprises. \citet{ueda2012enm} examines how the Bank of Japan (BOJ)'s unconventional monetary policy strategies affect the stock prices. He concludes that the strategies except for quantitative easing move the stock prices in the expected directions. In December 2010, the BOJ introduced the purchasing exchange-traded funds (ETFs) program and has been rapidly expanding the purchasing scale after April 2013. A few studies examine the effect on the stock markets because the BOJ is the only central bank purchasing stocks through ETFs. \citet{ueda2013ra} investigates the impacts of the ETFs purchasing by the BOJ and shows that the BOJ's unconventional monetary policies have positive effects on the TOPIX and the current exchange rates of the U.S. dollar against Japanese yen. \citet{harada2019boj} examine the cumulative treatment effects on the Nikkei 225 by the BOJ's EFTs purchasing program and show the effects on the Nikkei 225 have already reached around 20\% as of October 2017.

As mentioned above, economic policies often cause price distortion in stock markets and affect the market efficiency. We can consider that security prices are almost efficient in the long-run by rapidly reflecting new information on economic policies because market institutions and (information and communication) technologies are developed enough in the present day. However, when those were still developing, the prices did not always reflect new information on economic policies. In practice, \citet{hirayama2019mtv} shows that the degree of market efficiency in the weak sense changes and evolves over time using the prewar Japanese stock market data.\footnote{We call the period of prewar and wartime as ``prewar'' in this paper.} Therefore, it is reasonable to expect that we consider the market efficiency in the semi-strong sense also changes and evolves over time. Moreover, we can reveal the effect of policy changes on the financial markets under the closed economy because our dataset includes the wartime period. We pay attention to the transition process from the open economy to the closed economy in the prewar Japanese financial markets and characterize the period when the markets lost price formation function in the prewar Japanese financial system. We then focus on the relationship between the prewar Japanese financial markets and several government policy interventions to explore whether the semi-strong form market efficiency evolves over time.

Most of the previous studies employ the event study analysis to evaluate the semi-strong form EMH in accordance with \citet{ball1968eea} and \citet{fama1969asp}.\footnote{See \citet{brown1980msp} and \citet{mackinlay1997ese} for technical details.} The event study analysis is a very simple and convenient method; however, there is a serious problem. Particularly, we calculate a cumulative abnormal return (CAR) and test whether the CAR is statistically significantly different from zero on the specified event windows. We find that the event affects the stock prices when the CAR is statistically significantly different from zero. We can conclude the semi-strong form EMH is not supported. Note that the result of the event study analysis provides us estimates of only the short-run impact, not the long-run impact. Furthermore, the results of the event study analysis are quite unstable to even small changes in research design as shown in \citet{mcwilliams1997esm}. Hence, we adopt \citeapos{ito2014ism}{ito2014ism,ito2017aae} generalized least squares (GLS) based time-varying vector autoregressive (TV-VAR) model to investigate the long-run impacts of public information, especially government policy changes, in the stock markets. In the system of the GLS-based TV-VAR, we estimate the joint degree of market efficiency that is obtained from the time-varying impulse responses and evaluate how public information affects the financial market efficiency in the semi-strong sense. As a result, we can decide an argument about when the market lost price formation function in the prewar Japanese financial system.

This paper is organized as follows. Section \ref{semi_strong_sec2} provides a brief literature review that explains this paper's significance in finance and economic history. Section \ref{semi_strong_sec3} presents our empirical method for estimating the joint degree of market efficiency as the measurement of the semi-strong market efficiency that is based on the \citeapos{ito2014ism}{ito2014ism,ito2017aae} GLS-based TV-VAR model. Section \ref{semi_strong_sec4} presents our datasets that include new data of the performance index for the prewar Japanese stock and government bond markets estimated by \citet{hirayama2017rje,hirayama2017jsm,hirayama2018erj,hirayama2018jgb,hirayama2019lpp,hirayama2020jep,hirayama2021jgb} and presents some statistical test results. Section \ref{semi_strong_sec5} shows our empirical results using a GLS-based TV-VAR model and discusses the relationship between policy changes, major historical events, and time-varying semi-strong market efficiency in the prewar Japanese financial markets. Moreover, we discuss the period when the prewar Japanese financial markets lost the price formation mechanism. Section \ref{semi_strong_sec6} concludes the paper.

\section{Literature Review}\label{semi_strong_sec2}
This study examines \citetapos{fama1970ecm} semi-strong form EMH in the prewar Japanese financial market using \citeapos{ito2014ism}{ito2014ism,ito2017aae} GLS-based TV-VAR model. We expect that our study provides useful insights into the above-discussed important issues of finance and economic history. Previous research in the field of modern economic and financial history can be divided into two types: research on the economic system since the Meiji period, and research on the history of specific financial institutions and industries. The former focuses on policy changes and flow of funds. \citet{fujino2000qaj}, \citet{utsunomiya2013jfi}, and \citet{shibata2011mcw} systematically depict the structure of flow of funds. These studies also deal with the flow of funds among economic actors such as governments, households, corporations, and financial institutions. Typical examples of the latter research are \citet{kato1965bcw1,kato1965bcw2} and \citet{kasuya2006sir}, which clarify financial intermediation systems based on financial data of individual financial institutions. These approaches address changes in financial intermediation mechanisms or structures. However, as \citet{ito1995hsj} has pointed out, to grasp the actual condition of the financial intermediation system, it is necessary not only to analyze the structural changes described above, but also to take an approach to reexamine whether the functions of the system are being fully utilized. This is because improving the financial structure does not necessarily mean an efficient financial intermediation function. For example, in the financial system in developing East Asia in 1990, the financial structure and its function diverged. In this study, we focus on this point and examine the efficiency of financial markets to clarify whether financial intermediation is functioning using market indicators showing accurate investment performance from 1924 to 1945. In this study, we focus on the time-varying nature of the semi-strong form market efficiency to quantitatively grasp the functions of the financial markets in prewar Japan, which will contribute to complementing conventional studies of economic history. There have been few previous studies on the functioning of stock markets. In previous studies, economic policy by Minister of Finance Korekiyo Takahashi from 1932 to 1936 has been one of the main themes of policy analysis in economic history. Many of them believe that the recovery from the global Great Recession became possible by policy packages such as fiscal expansion policies, currency depreciation, and monetary easing policies. However, quantitative studies that have analyzed the effects of this policy are limited to \citet{shibamoto2014era}. They explore that this policy had a large effect on raising expected inflation before seceding from the gold standard system.

Studies on financial markets in the early 1930s have focused on the government bond market, but not on the stock market. The main monetary policy at that time was focused on the government bond market and the foreign exchange market. The interest of policymakers in the stock market was low. In 1932, however, stock prices rose significantly and the situation in Japan was different from the stagnation of stock indices in Europe and the U.S. Therefore, it is necessary to reexamine the situation quantitatively. In this sense, \citet{bassino2015iet}, a financing study on the stock market before and during the war, is a valuable contemporary study. However, the market data used (daily data in the {\it{Economic Yearbook of the Toyo Keizai}} ({\it Oriental Economist}) will need to be reconsidered. First, the stock price data they use do not consider that correction has been made to the settlement of rights and additional payments, or the return from dividends.\footnote{See Appendix for details.} Second, Japanese government bond (JGB) data are limited to 5\% coupon bonds and do not consider changes in spreads by interest rate, nor do they include 3.5\% coupon bonds issued in large volumes after the outbreak of the Second Sino-Japanese War.\footnote{Additionally, income returns, which account for the majority of total JGB returns, are not considered.} Since the restoration of the gold embargo, as can be seen from the examples of the BOJ's underwriting of government bonds and the revision of the Capital Flight Prevention Law, Japanese government intervention was strengthened in the government bond and foreign exchange markets earlier than in the stock market. By contrast, in the stock market, strong government intervention such as the Share Price Control Ordinance and Stock Price Support Organizations was not implemented until the wartime. Quantitative analysis of market efficiency will help to identify when government price intervention in the early Showa period had a significant impact on the market. This is because policy effects cannot be sufficiently measured only by studying the history of economic policy at that time and by checking trends inflow of funds. Quantitative assessment of market impact will also be an effective tool for studying economic history.

From a different perspective the three main issues related to market functions in research on economic history are as follows. First, there is a view that the stock market stagnated because investors dropped out after the Showa Financial Depression in 1927 and that the stagnation of the market continued even after the World War I\hspace{-.1em}I (WWI\hspace{-.1em}I). Another view states that although the price control of the government bond market was strengthened, the degree of freedom in price fluctuations in the stock market was secured to some extent. The two views on the stock market could be explored to a certain extent through the time-series evolution of market volatility and comparisons of other assets. \citet{hirayama2018erj} provides descriptive statistics such as the stock market price volatility in the U.S. and Japan, and compares the difference between the markets. Based on the index he calculated, the risk level of the stock market in the 1940s (until November 1944) was 11.8\% in Japan and 13.5\% in the U.S. Although the risk level of the Japanese stock market is relatively low, it can be said that a certain degree of volatility has been maintained compared with the U.S. and Japan. Additionally, the yield spread, which is the difference between the yield on the long-term government bonds and the equity earnings yield, has been rising in both markets, while the relative positions have changed alternately since the late 1930s. In terms of the comparison between Japan and the U.S. in the secondary market, it may not be said that only the Japanese market has shrunk. In this way, it is possible to supplement the view of economic history by quantitative bilateral market analysis. Second, whether corporate finance functioned mainly in the securities market or in the traditional banking industry before and after World War I (WWI) remains debatable. Although the views of these discussions do not agree, a quantitative understanding of changes in the efficiency of the stock market and other markets will have certain implications for the discussion. This will contribute to broaden the scope of this discussion by examining the functioning of securities markets. 

Some studies examine the efficient market hypothesis using the prewar Japanese stock market data. \citet{kataoka2004b} investigate whether the prewar Japanese stock market was efficient using the daily stock prices in 1900 alone. They find that the market is efficient in the weak sense but the market is not efficient in the semi-strong sense. \citet{suzuki2012pwt} assumes known breakpoints and employs the \citetapos{choudhry2010ww2} method to examine the semi-strong EMH using the Japanese daily stock market data during WWI\hspace{-.1em}I. He finds that market efficiency declined after the start of the war and that the known breakpoints are consistent with the variation of stock prices. \citet{bassino2015iet} explore the weak-form EMH in the 1930s' Japanese stock market using the generalized autoregressive conditional heteroskedasticity (GARCH)-in-mean model and argue that the market deviated from weak-form efficiency. \citet{hirayama2019mtv} employs the long time-series data from 1878 to 1945 and measures the time-varying degree of market efficiency in the weak sense to test the \citetapos{lo2004amh} AMH. He finds that the prewar Japanese stock market is almost efficient and the degree of market efficiency in the prewar Japanese stock market varied with time and that its variations corresponded with major historical events. \citet{suzuki2019gke} apply the unit root test of \citet{perron1989tgc} with a known breakpoint to examine whether stock prices formed efficiently after the Kanto Great Earthquake. They find that stock prices reflected various data correctly even after the earthquake occurred and conclude the semi-strong form EMH is supported. Therefore, whether the semi-strong form EMH is supported in the prewar Japanese stock market is controversial. We then focus on the time-varying structure of the stock market and dynamically observe changes in semi-strong form efficiency, after which we should be able to see the transition of market functions in detail.

Lastly, there is a view that although the primary market for stocks in the prewar and wartime periods did not function, the secondary market was bloated and speculative. As pointed out by \citet[p.235]{kobayashi2012hjs}, it is important to consider whether the prewar market was functioning properly by distinguishing between primary and secondary markets. For the primary market, structural analysis such as institutional changes and flow of funds will be effective; however, for the secondary market, it will be necessary to check for speculative movements. Particularly, questions such as whether currency speculation arising from the restoration of the gold embargo, as described in detail by \citet{fukai1941r70}, actually took place in the secondary stock market (synchronicity between markets), and whether the secondary stock market during the war was affected by external factors such as the war situation (wartime price fluctuations) will be the subjects of historical research. From the above perspective, the testing the semi-strong form EMH in the prewar Japanese stock market may provide some suggestions on the prewar financial intermediation function in economic history.

\section{The Method}\label{semi_strong_sec3}
We first introduce \citeapos{ito2014ism}{ito2014ism,ito2017aae} GLS-based TV-VAR model to analyze the time-varying nature of the semi-strong market efficiency on prewar Japanese financial markets. We suppose that ${\ve{p}_t}$ is a price vector of the three securities (stock, government bond, and foreign exchange rate) in $t$ period. Our main focus is reduced to the following condition: 
\begin{equation}
\ex\left[ {\ve{x}_{t}}\mid\mathcal{I}_{t-1}\right]=\ve{0}, \label{eq1}
\end{equation}%
where ${\ve{x}_{t}}$ denotes a return vector of the securities in $t$ period. We calculate that the $i$-th element of ${\ve{x}_{t}}$ is $\ln{p_{i,t}}-\ln{p_{i,t-1}}$ for each security. Note that Equation (\ref{eq1}) implies that all conditional expected returns of the three securities at $t$ period given the information set available in $t-1$ period are zero. If $\ve{x}_{t}$ is stationary, we denote the time-series process of $\ve{x}_{t}$ using the Wold decomposition as
\begin{eqnarray}
\ve{x}_{t}&=&\ve\mu+\Phi_{0}\ve{u}_{t}+\Phi_{1}\ve{u}_{t-1}+\Phi_{2}\ve{u}_{t-2}+\cdots\nonumber\\
&=&\ve\mu+\Phi(L)\ve{u}_t, \label{eq2}
\end{eqnarray}
where $\Phi(L)$ is a matrix lag polynomial of a lag operator, $\mu$ is the mean of $\ve{x}_t$, and $\ve{u}_t$ follows an independent and identically distributed multivariate process with a mean of zero vector. We assume that coefficient matrices $\{\Phi_i\}^\infty_{i=0}$ are $k\times k$ dimensional parameter matrices. We then compute the impulse-response functions along with the identification assumptions such as $\Phi_0=I$. Note that the EMH holds if and only if $\Phi_i=0$ for all $i>0$, which suggests that the market deviation from the efficient market reflects the impulse response, a series of $\ve{u}_t$. In this study, we construct a relative degree based on the impulse response to investigate whether the semi-strong form EMH holds for the prewar Japanese financial markets. 

We can obtain that the impulse response is to employ a VAR model and to algebraically compute its coefficient estimates. Suppose that the vector return process $\ve{x}_t$ of the three securities is invertible under some assumptions. Then, we consider the following standard VAR($q$) model:
\begin{equation}
\ve{x}_{t}=\ve\nu+A_{1}\ve{x}_{t-1}+A_{2}\ve{x}_{t-2}+\cdots+A_{q}\ve{x}_{t-q}+\ve{\varepsilon}_{t}; \ \ t=1,2,\ldots,T, \label{eq3}
\end{equation}
where $\ve\nu$ is a vector of intercepts; $\ve{\varepsilon}_{t}$ is a multivariate error term with $\ex\left[\ve{\varepsilon}_{t}\right]=\ve{0}$, \ $\ex\left[\ve{\varepsilon}_{t}^{2}\right]=\sigma_{\ve{\varepsilon}}^{2} I $, and $\ex\left[\ve{\varepsilon}_{t}\ve{\varepsilon}_{t-m}\right]=\ve{0}$ for all $m\neq 0$. We can measure a relative degree about the semi-strong form market efficiency that varies over time according to the following procedure. First, we compute a cumulative sum of the coefficient matrices of the impulse response:  
\begin{equation}
\Phi(1)=\left(I - A_1 - A_2 - \cdots - A_q\right)^{-1},\label{eq4}
\end{equation}
We next define a joint degree of market efficiency as the measurement of the semi-strong form market efficiency:
\begin{equation}
\zeta=\sqrt{\mbox{max} \left[(\Phi(1) - I)'(\Phi(1) - I)\right]},\label{eq5}
\end{equation} 
to measure the deviation from the efficient market. We can understand that in the case of the efficient market where $A_1=A_2=\cdots=A_q=0$, the degree $\zeta$ becomes zero; otherwise, $\zeta$ deviates from zero. Thus, we call $\zeta$ the joint degree of market efficiency. When we find a large deviation of $\zeta$ from $0$ (both positive and negative), we can regard some deviation from one as evidence of market inefficiency. Furthermore, we can construct this degree that would vary over time when we obtain time-varying estimates of the coefficients in Equation (\ref{eq3}).

We estimate TV-VAR coefficients in each period to obtain the degree defined in Equation (\ref{eq5}) in each period. In practice, we employ a model in which all the VAR coefficients, except for the one that corresponds to intercept terms, $\ve\nu$, follow independent random walk processes. In other words, we assume:
\begin{equation}
A_{l,t}=A _{l,t-1}+V_{l,t}, \ \ (l=1,2,\cdots,q), \label{eq6}
\end{equation}
where $\{V_{l,t}\}$ is a $q\times t$ dimensional error term matrix. We assume that the matrix satisfies $E\left[V_{l,t}\right]=\ve{O}$ for all $t$, \ $E\left[vec(V_{l,t})'vec(V_{l,t})\right]=\sigma_v^{2} I$ and $E\left[vec(V_{l,t})'vec(V_{l,t-m})\right]=\ve{O}$ for all $l$ and $m\neq 0$. \citeapos{ito2014ism}{ito2014ism,ito2017aae} method allows us to estimate the TV-VAR model:
\begin{equation}
\ve{x}_{t}=\ve\nu+A_{1,t}\ve{x}_{t-1}+A_{2,t}\ve{x}_{t-2}+\cdots+A_{q,t}\ve{x}_{t-q}+\ve\varepsilon_{t},  \label{eq7}
\end{equation}
together with Equation (\ref{eq6}). 

To conduct statistical inference on a joint degree of market efficiency in the semi-strong sense, we apply a residual bootstrap technique to the TV-VAR model above. In practice, we build a set of bootstrap samples of the TV-VAR estimates under the hypothesis that all the TV-VAR coefficients are zero. This procedure provides us with a (simulated) distribution of the estimated TV-VAR coefficients, assuming the three securities return processes are generated under the semi-strong form EMH. Thus, we can compute the corresponding distribution of the impulse response and degree of market efficiency in the semi-strong sense. Finally, by using confidence bands derived from such simulated distribution, we conduct statistical inference on our estimates and detect periods when the prewar Japanese financial markets experienced inefficiency.

\section{Data}\label{semi_strong_sec4}

This paper studies the dynamics of the semi-strong efficiency of the prewar Japanese financial markets, investigating how the security prices (or total returns) responded to policy changes and external events. That is, we consider not only the stock but also the government bond and the foreign exchange when we measure the time-varying joint degree of market efficiency in the prewar Japanese stock market. As stock prices, we utilize the total return of the equity performance index (EQPI) that are calculated by \citet{hirayama2017rje,hirayama2017jsm,hirayama2018erj,hirayama2019lpp,hirayama2020jep}. Note that the EQPI is the first and only capitalization-weighted index in the prewa Japanese stock market. Then the government bond performance index (GBPI) of \citet{hirayama2018jgb,hirayama2021jgb} is used as the government bond prices. Hirayama calculates the total return for the prewar Japanese government bond and constructed GBPI as a new government bond price index. Lastly, we obtain current exchange rates of the U.S. dollar against Japanese yen from the {\it{Bulletin of Financial Information}} (Ministry of Finance, Japan) and \citet{boj1947hsw} as the foreign exchange prices.\footnote{Please see the appendix at the end of this paper on how the dataset is constructed.} 

The sample periods of all securities are both from June 1924 to August 1945. We take the log first difference of the time series of prices to obtain the returns of the indices.
\begin{center}
(Table \ref{semi_strong_table1} around here)
\end{center}
Table \ref{semi_strong_table1} shows the descriptive statistics for the returns. We confirm that the mean (standard deviation) of returns on the EQPI is higher (lower) than that of the other securities. This means that the EQPI was considered a relatively risky asset in prewar Japan. We also confirm that the standard deviation of the returns on GBPI is the lowest in all of the securities. This implies that the government bond was considered a relatively low-risk asset in prewar Japan.

Table \ref{semi_strong_table1} also shows the results of the unit root test with descriptive statistics for the data. For estimations, all variables that appear in the moment conditions should be stationary. We apply the \citetapos{elliott1996eta} augmented Dickey--Fuller generalized least squares (ADF-GLS) test to confirm whether the variables satisfy the stationarity condition. We employ the modified Bayesian information criterion (MBIC) instead of the modified Akaike information criterion (MAIC) to select the optimal lag length. This is because from the estimated coefficient of the detrended series, $\hat\psi$, we do not find the possibility of size-distortions (see \citet{elliott1996eta}; \citet{ng2001lls}). The ADF-GLS test rejects the null hypothesis that the variables (all returns) contain a unit root at the 5\% significance level.

\section{Empirical Results}\label{semi_strong_sec5}

\subsection{Preliminary Estimation}
We first assume a time-invariant VAR($q$) model with constant parameters and employ \citetapos{schwarz1978edm} Bayesian information criteria to select the optimal lag order in our preliminary estimation. Table \ref{semi_strong_table1} summarizes our preliminary results for a time-invariant VAR($q$) model using the whole sample. In the estimation, we choose first-order vector autoregressive (VAR($1$)) model for the estimation. 
\begin{center}
(Table \ref{semi_strong_table2} around here)
\end{center}
Table \ref{semi_strong_table2} shows that (1) autoregressive estimates in the VAR model for the EQPI and Exchange are almost similar, and (2) the autoregressive estimates of the GBPI is larger than those of EQPI and Exchange. This means that the government bond market is the most inefficient in the prewar Japanese financial market. Note that because of this result as well as the limited explanatory power of the time-invariant VAR model, we should pay more attention to the time-varying nature of the semi-strong market efficiency on the prewar Japanese financial markets.

We employ \citetapos{hansen1992a} parameter constancy test under the random parameters hypothesis to detect whether the parameters are constant in the above VAR($q$) model. Table \ref{semi_strong_table2} also presents the test statistics; we reject the null of constant parameters against the parameter variation as a random walk at the 5\% significance level. Therefore, we estimate the time-varying parameters of the above VAR model to investigate whether gradual or rapid changes occur in the prewar Japanese financial markets. These results suggest that the time-invariant VAR($q$) model is not suitable to explain our data and that the TV-VAR($q$) model is a better fit.

From a historical viewpoint, the prewar Japanese economy experienced various exogenous shocks such as bubbles, economic or political crises, policy changes, and wars. Table \ref{semi_strong_table3} summarizes the major historical events in the prewar Japanese economy.
\begin{center}
(Table \ref{semi_strong_table3} around here)
\end{center}
We consider that the events affected the stock price formation. Accordingly, we estimate the degree of market efficiency using the GLS-based TV-VAR model in the next subsection.

\subsection{The Changes in the Financial Market Functions}
\begin{figure}[tbp]
 \caption{Time-Varying Degree of Market Efficiency (Joint)}
 \label{semi_strong_fig1}
 \begin{center}
 \includegraphics[scale=0.8]{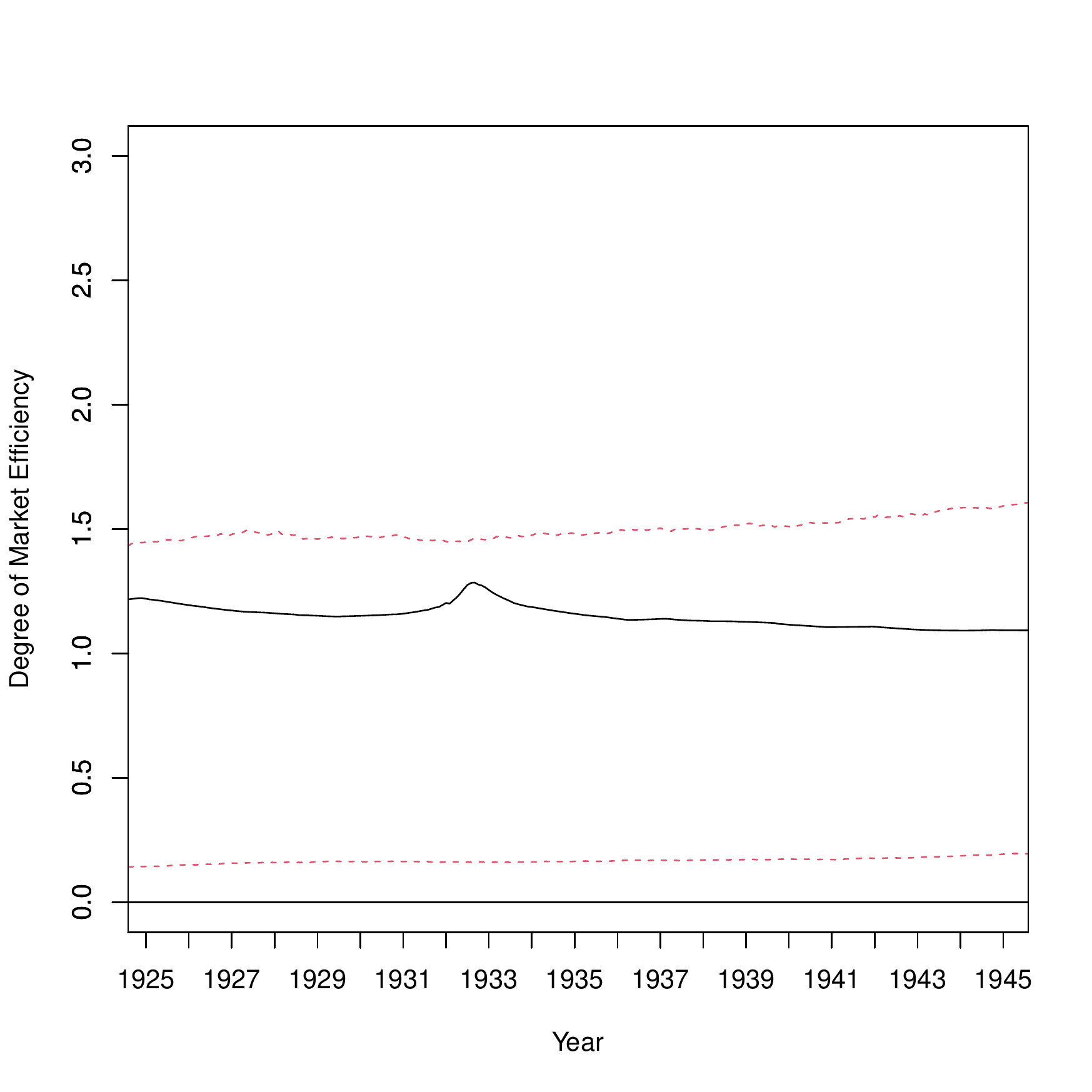}
\vspace*{5pt}
{
\begin{minipage}{420pt}
\underline{Notes}:
\begin{itemize}
 \item[(1)] The dashed red lines represent the 95\% confidence intervals of the efficient market degrees.
 \item[(2)] We run the bootstrap sampling 5,000 times to calculate the confidence intervals.
 \item[(3)] R version 4.1.0 was used to compute the estimates.
\end{itemize}
\end{minipage}}%
\end{center}
\end{figure}

\begin{figure}[tbp]
 \caption{Time-Varying Degree of Market Efficiency (Individuals)}
 \label{semi_strong_fig2}
 \begin{center}
 \includegraphics[scale=0.4]{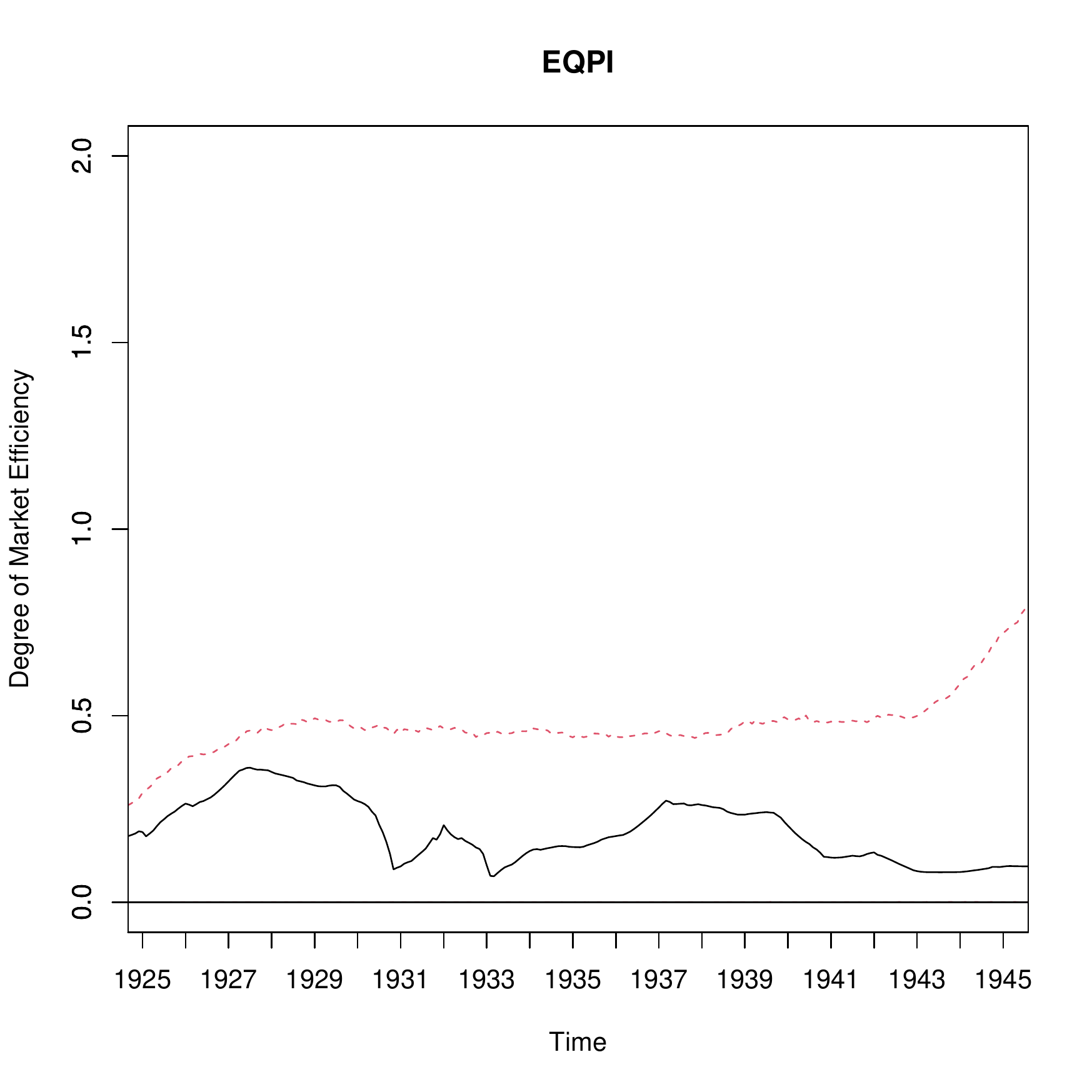}
 \includegraphics[scale=0.4]{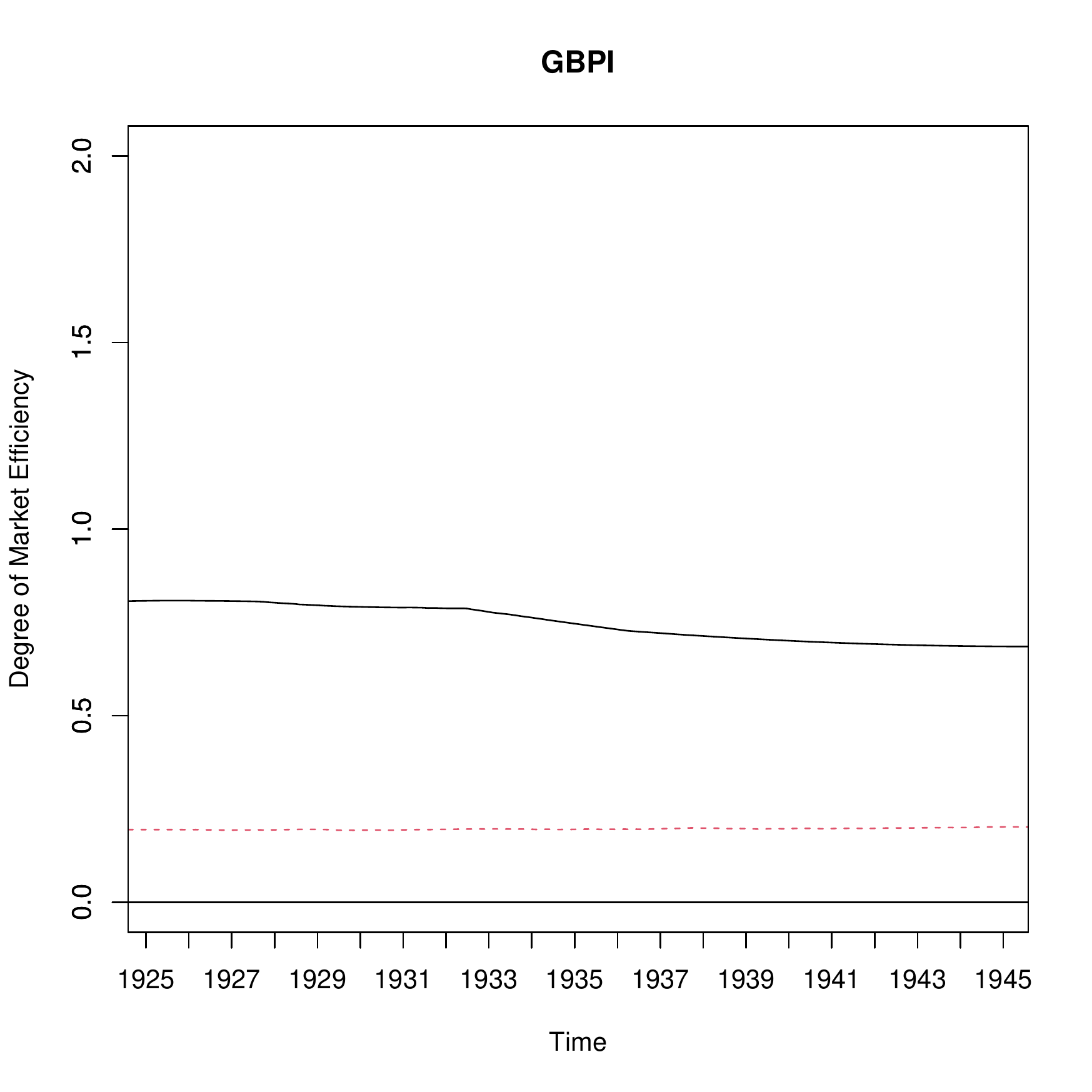}
 \includegraphics[scale=0.4]{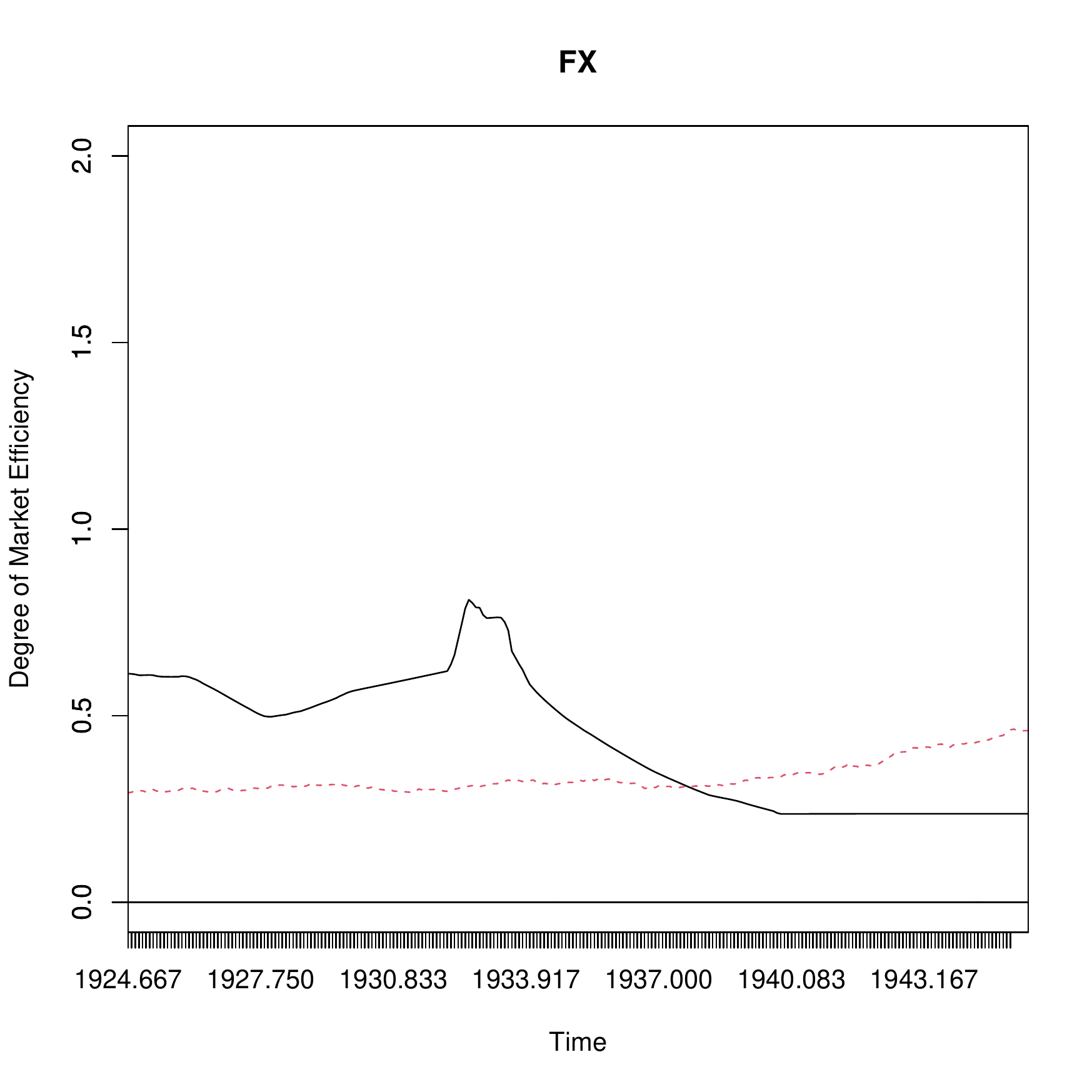}
\vspace*{5pt}
{
\begin{minipage}{420pt}
\underline{Notes}:
\begin{itemize}
 \item[(1)] The dashed red lines represent the 95\% confidence intervals of the efficient market degrees.
 \item[(2)] We run the bootstrap sampling 5,000 times to calculate the confidence intervals.
 \item[(3)] R version 4.1.0 was used to compute the estimates.
\end{itemize}
\end{minipage}}%
\end{center}
\end{figure}

\begin{center}
(Figure \ref{semi_strong_fig1} around here)
\end{center}
Figure \ref{semi_strong_fig1} shows that the joint degree of market efficiency in the prewar Japanese financial markets fluctuated over time.\footnote{Note that are some related studies about examining the weak-form EMH in prewar Japan such as \citet{ito2016meg,ito2018fpe}. They estimate the time-varying degree of market efficiency in the prewar Japanese rice markets and find that the market efficiency changes through time.} We confirm that the prewar Japanese financial markets are almost efficient in the sense of the semi-strong form over time entire prewar period, and the effects of exogenous shocks such as economic policy changes and wars are quickly reflected in prices. However, the changes in the joint degree show that the efficiency deteriorated until July 1932 and then improved to the level before the deterioration, suggesting that the joint degree of market efficiency fluctuates over time. In the following, we measure the individual degree of market efficiency for each market: stocks, government bonds, and foreign exchanges to explore the time-varying factors of the price determination function from a historical perspective.\footnote{When stock price data of the Tokyo Stock Exchange, a major stock in the prewar period, is used in the verification of the stock market, there is a drawback that it is highly individualized because the performance of a specific industry (financial sector) is strongly reflected. The EQPI is used in this study because an indicator containing a large number of stock data (all possible stocks) should be used to measure the efficiency of the stock market as a whole. Similarly, in the government bond market, the difference in return was largely due to differences in coupons in the prewar period, so government bond indices covering all types of issues were used (See \citet{hirayama2018erj,hirayama2018jgb} for details.).}
\begin{center}
(Figure \ref{semi_strong_fig2} around here)
\end{center}
Figure \ref{semi_strong_fig2} shows the individual market efficiency in stock, government bonds, and foreign exchange markets changes over time. First, while the stock market was efficient, the government bond market was inefficient over time. This is considered to be because the market mechanism, in which changes in the external environment and other factors are reflected in prices, worked in the stock market over time. Second, the market mechanism did not work in the government bond market, as the autocorrelation of returns was extremely high in the market over time.\footnote{The first-order autocorrelation of the return data of the Japanese government bonds used in this study reaches 0.43.} Last, although the foreign exchange market was inefficient until the early 1930s, market efficiency has been improving since April 1933, when the U.S. and Japan withdrew from the gold standard system. In addition to the above time-varying efficiencies of each market, the changes will be discussed in detail below.

First, efficiency in the stock market changes in response to changes in external events and environments. After the Showa financial crisis in 1927 and the shift to the gold standard system in January 1930, the stock market was on a downward trend, but market efficiency improved until November 1930. However, when the economic situation deteriorated after the gold ban was lifted, speculation grew that the gold standard would be withdrawn. As a result, speculative moves to expect a rise in stock prices increased, and market volatility rose rapidly and stock market efficiency declined. When the decision to withdraw from the gold standard system was made in December 1931, the stock market was on an upward trend amid the rapid depreciation of the yen against the U.S. dollar. At this stage, the uncertainty over changes in the exchange rate regime was resolved, and efficiency recovered. After the decision by the United States to withdraw from the gold standard system was made in April 1933, efficiency declined, but it increased in the process of strengthening the wartime regime after the outbreak of the second Sino-Japanese War in 1937.

Second, the efficiency of the government bond market changes slightly in July 1932. This is considered to have been influenced by the following policy changes. The decline in government bond prices in 1931 increased unrealized losses on government bonds held by financial institutions. Therefore, in order for financial institutions to continue to purchase government bonds in the future, the government needed to increase incentives for financial institutions to hold government bonds. The Minister of Finance Korekiyo Takahashi established the ``Act on Valuation of Government Bonds'' in July 1932. This law sets the minimum level of the appraised value of government bonds held by financial institutions as the issue price so that no appraisal loss occurs in accordance with changes in market prices. In November 1932, the Bank of Japan began underwriting government bonds, and the government strengthened its policy of supporting the prices, thus suppressing price volatility in the government bond market. As a result, the situation in which the autocorrelation of the return on investment in the government bonds was extremely high continued. Although price volatility disappeared during the war period, pricing in the market is considered inefficient in the sense of \citet{fama1970ecm} if autocorrelation is extremely strong.\footnote{\citet{shiller1981dsp} argues that the efficient market hypothesis may be rejected if there exists extreme volatility. However, the existence of extreme volatility is not a sufficient condition to reject the martingale property based on the efficient market hypothesis. Therefore, the market is inefficient in the sense of \citet{fama1970ecm} when autocorrelation is extremely strong even if the volatility of return (or price) is low.} In other words, it can be said that, against the background of the thorough government bond management policy, the profitability of the government bond market has become extremely highly autocorrelated and inefficient over time.

Third, the efficiency of the foreign exchange market has been reversed before and after the changes to the two foreign exchange policies (the ``Capital Flight Prevention Law'' enacted in July 1932 and the ``Foreign Exchange Control Law'' enacted in May 1933). Market efficiency declined before the enactment of the Capital Evacuation Prevention Law, but after the enactment of the Foreign Exchange Control Law, efficiency improved and inefficiency was eliminated. Before the enactment of the two laws, the differences in the responses of each country to the gold standard system and the change in expectations for it were mixed. The gold standard system is a mechanism in which economic differences among countries are adjusted through the exchange system, and exchange rates do not reflect macroeconomic trends. Therefore, exchange rate fluctuations do not fully reflect the external environment and are considered highly inefficient. This is because the more the exchange rate determined by the government deviates from the economic situation, the more active the speculative transactions are in anticipation of policy changes, such as a change in the exchange rate or withdrawal from the gold standard. The adoption and departure of the gold standard system entail large fluctuations in exchange rates and therefore represents a golden opportunity for market participants. \citet[p.245]{fukai1941r70} shows that under the gold standard system, there were quite a few transactions in which foreign currencies were bought at low prices in the hope of being withdrawn in the future, and then sold after the ban was re-imposed to make profits. 

After the return to the gold standard system, foreign currency purchases occurred immediately after the return, but these were transactions to repurchase foreign currency sales in the hope of the return to the gold standard system. Indeed, the turmoil surrounding the gold standard system from 1929 to 1933 stimulated speculation in exchange markets. After the enactment of the two foreign exchange laws, major countries were freed from the spell of the gold standard system, and speculative buying of dollars subsided, hovering in the range from 3.3 to 3.7 Japanese yen to the U.S. dollar, and the rate of market volatility also declined, making foreign exchange markets more efficient. In October 1939, when Japan's reference currency for foreign exchange was changed from the pound to the U.S. dollar, due to the outbreak of the war in Europe, the exchange rate was revised to approximately 4.27 Japanese yen to the U.S. dollar, and in January 1942, the Minister of Finance announced that the exchange rate was officially fixed at 4.25 Japanese yen to the U.S. dollar. During this period, the efficiency of the foreign exchange market was maintained.

As described above, we find that (1) the level of market efficiency is significantly different in each market, and (2) the efficiency of the stock market and the foreign exchange market, excluding the government bond market, is changing due to market policies of the government, and it is highly likely that the adaptive market hypothesis has been supported. We confirm that the effect of ``Act on Valuation of Government Bonds'' in July 1932 was detected as mentioned in \citet{hirayama2018jgb} and the efficiency was low not only in the wartime period but also in the entire sample period due to the extreme autocorrelation strength. In addition, it was newly detected that the function of price determination in the market had been lost even before the strengthening of financial control, and it was confirmed that government bond management policy had been thoroughly implemented to a considerable extent even before the implementation of monetary and fiscal policy by the Minister of Finance Korekiyo Takahashi. We conclude that semi-strong efficiency was maintained in the prewar Japanese financial markets throughout the time period in this study is consistent with \citet{hirayama2018erj,hirayama2018jgb} on the stock market that the market function was maintained during wartime.

Moreover, looking at the financial markets as a whole, although the joint degree of market efficiency was secured, it was confirmed that the efficiency declined in July 1932, several months after the departure from the gold standard system over time. From November 1930 to December 1931, efficiency declined due to rising price volatility in the stock market, but this did not affect the joint degree of market efficiency as a whole. On the other hand, the sharp rise in volatility and inefficiency in foreign exchange markets from January 1932 to July 1932 affected the joint degree of market efficiency as a whole. In other words, during the period from 1924 to 1945, the major changes in the market functions of financial markets as a whole were not due to the fall of the New York stock market or to financial control during wartime, but to the turmoil in the exchange markets after the departure from the gold standard system. This does not deny the conventional view that pre- and post-war financial markets have become discontinuous since the end of World War II, but it can be said that the pricing mechanism in Japanese financial markets had reached a major turning point in July 1932, when the ``Capital Flight Prevention Law'' and ``Act on Valuation of Government Bonds'' were enacted. 

\subsection{Time-Varying Impulse Responses}
\begin{figure}[tbp]
 \caption{Time-Varying Impulse Responses}
 \label{semi_strong_fig3}
 \begin{center}
 \includegraphics[scale=0.9]{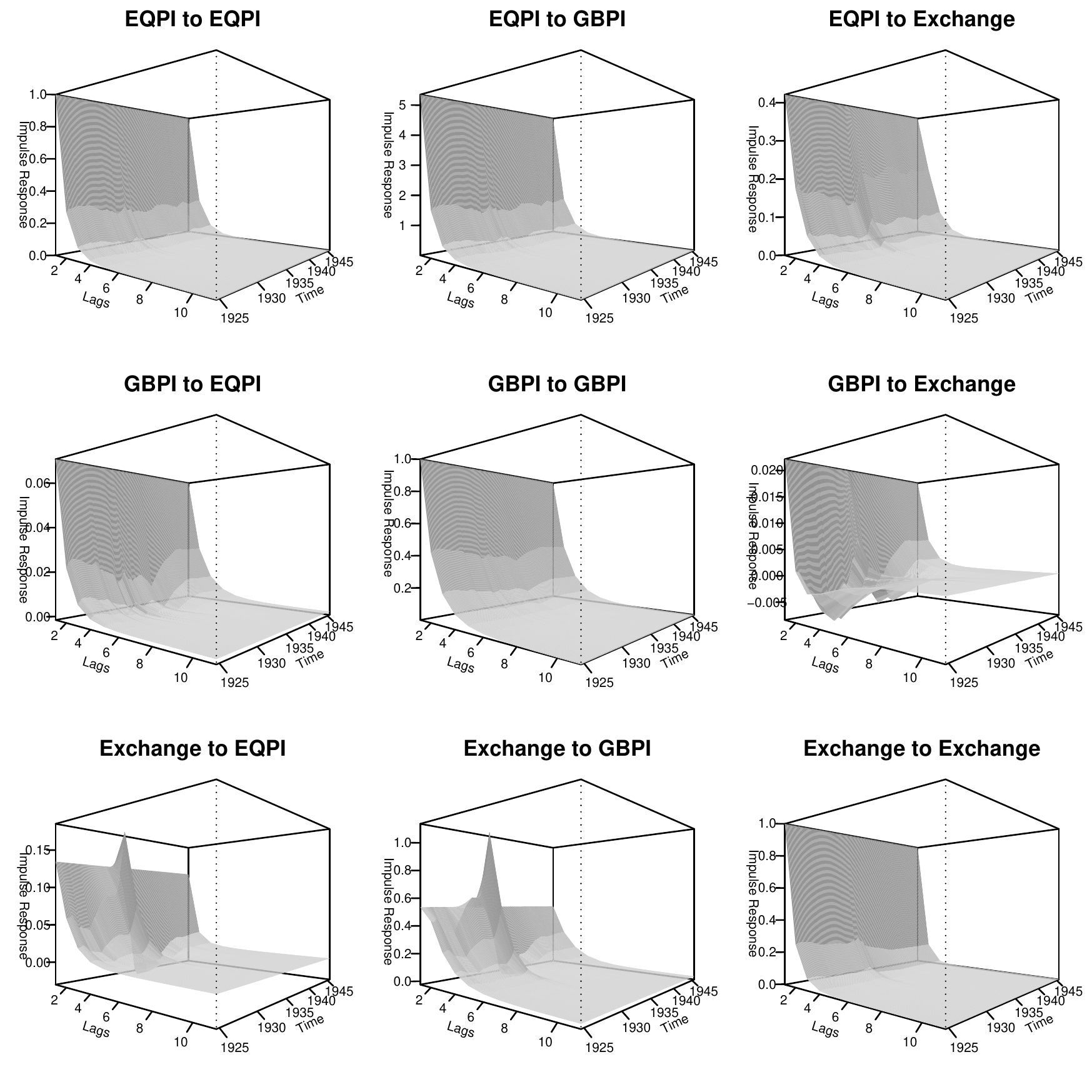}
\vspace*{5pt}
{
\begin{minipage}{420pt}
\underline{Note}: R version 4.1.0 was used to compute the estimates.
\end{minipage}}%
\end{center}
\end{figure}

\begin{figure}[tbp]
 \caption{Static Impulse Responses}
 \label{semi_strong_fig4}
 \begin{center}
 \includegraphics[scale=0.8]{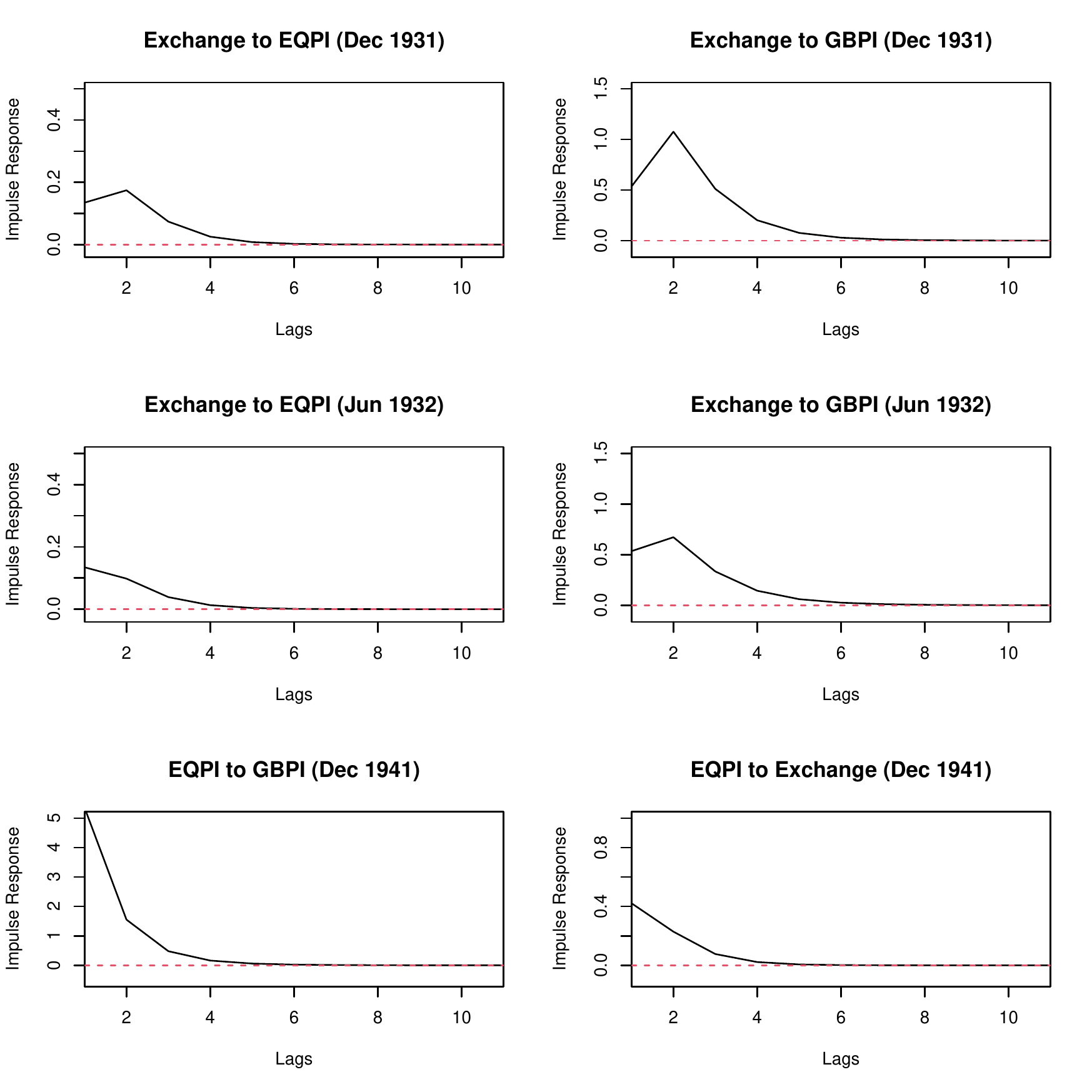}
\vspace*{5pt}
{
\begin{minipage}{420pt}
\underline{Note}: R version 4.1.0 was used to compute the estimates.
\end{minipage}}%
\end{center}
\end{figure}

Next, we examine the time-varying impulse responses to investigate how external events as shocks affect the three markets.
\begin{center}
(Figure \ref{semi_strong_fig3} around here)
\end{center}
Figure \ref{semi_strong_fig3} shows the time-varying impulse responses of each market, while the non-diagonal figures show that shocks in the foreign exchange market are the greatest, followed by shocks in government bonds and stocks. In particular, the enactment of the capital flight prevention law in July 1932 seems to have affected stocks and government bonds through the exchange rate. However, it can be confirmed that the outbreak of the Pacific War in December 1941 had little impact on each market. Next, in order to examine the effect of the event in detail, let us check the static impulse response at a specific point in time by dividing the above time-varying impulse responses. In particular, attention should be paid to the following three points: (1) the abandonment of the gold standard system in December 1931, (2) the enactment of the Capital Escape Prevention Law in July 1932, and (3) the start of the Pacific War in December 1941.
\begin{center}
(Figure \ref{semi_strong_fig4} around here)
\end{center}
Figure \ref{semi_strong_fig4} shows the static impulse response at the three points. First, it can be confirmed that the shock caused by the outbreak of the Pacific War diminished more rapidly than the shock caused by the abandonment of the gold standard system and the enactment of the Capital Escape Prevention Law. This means that the market reflected information on external events quickly over time. Furthermore, it became clear that the static impulse response of the outbreak of the Pacific War was incomparably smaller than that of the change in foreign exchange policy. These results are consistent with \citet[pp.271--278]{ito1989jof} which shows that despite the decision to ban gold re-exports in December 1931 and the enactment of the Capital Escape Prevention Law in July 1932, there was a time lag before strict exchange controls were implemented. Therefore, during the period before Japan's policy stance on foreign exchange became thoroughly controlled, speculation in the foreign exchange market intensified, and market efficiency declined.

\section{Concluding Remarks}\label{semi_strong_sec6}

It is believed that the Japanese financial system was financial market-centered before the end of WWI\hspace{-.1em}I, and there has been no consensus on when the Japanese financial system would change to be bank-centered after the end of WWI\hspace{-.1em}I. This paper examines when the financial markets lost price formation function in prewar Japan in the sense of \citetapos{fama1970ecm} semi-strong form market efficiency using a new dataset obtained from \citet{hirayama2017rje,hirayama2017jsm,hirayama2018erj,hirayama2018jgb,hirayama2019lpp,hirayama2020jep,hirayama2021jgb}. Particularly, we focus on the relationship between the prewar Japanese financial markets and several government policy interventions to explore whether the semi-strong form market efficiency evolves over time. We consider that government policy interventions change the semi-strong market efficiency through the distortion of security prices such as stock, government bonds, and foreign exchange. To capture the long-run impact of government policy interventions against stock markets, we measure the time-varying joint degree of market efficiency and the time-varying impulse responses based on \citeapos{ito2014ism}{ito2014ism,ito2017aae} GLS-based TV-VAR model instead of the conventional approach such as the event study analysis used in most of the previous studies. 

We summarize the results as follows. First, the joint degree of market efficiency in the prewar Japanese financial markets fluctuated over time because of external events such as policy changes and wars. These results are consistent with the results of \citet{hirayama2019mtv} who measures the degree of market efficiency in the sense of weak-form using the prewar Japanese stock market data. Second, the semi-strong form EMH is almost supported in the prewar Japanese financial markets as well as the results of \citet{suzuki2019gke}. They show that stock prices reflected various data correctly even after the Kanto Great Earthquake occurred; and they conclude the semi-strong form EMH is supported. Third, we can see that the impulse responses in the case of the outbreak of the Pacific War decay more rapidly than those of the abandonment of the gold standard and the establishment of the foreign exchange control law. It means that \citetapos{lo2004amh} AMH is also supported in the prewar Japanese financial markets even if we consider that the public information affects the markets because the markets now reflect the information of the external events to the security prices faster than before. Last, we identify the point in time at which the price formation function of the Japanese financial markets lost. As a result, it became clear that significant changes in the function occurred in 1932, particularly in the foreign exchange market, rather than in the stock market. Therefore, we conclude that 1932 was a turning point in terms of the price formation function in the prewar Japanese financial markets.



\section*{Acknowledgments}

The authors would like to thank Mikio Ito, Yumiko Miwa, Masato Shizume, Shiba Suzuki, Tatsuma Wada, Takenobu Yuki, and the seminar participants at Keio University for their helpful comments and suggestions. The author (Noda) is also grateful for the financial assistance provided by the Japan Society for the Promotion of Science Grant in Aid for Scientific Research, under grant numbers 17K03809, 18K01734, and 19K13747. All data and programs used are available upon request.

\clearpage


\setcounter{section}{0}
\setcounter{subsection}{0}

\renewcommand{\thesection}{A.\arabic{section}}
\renewcommand{\thesubsection}{A.\arabic{subsection}}

\section*{Appendix}
 
\subsection{Dataset on the Government Bonds Market}
In the past, the yields on the 5\% Loan (Mark Ko) and 4\% Loan (1st issue) were used as indicators of the prewar and wartime government bond markets. However, analysis of these individual issues does not show actual investment performance in the prewar and wartime Japanese government bond (JGB) markets, because yield spreads exist due to differences in coupon rates. \citet{hirayama2018jgb,hirayama2021jgb} provides a capitalization-weighted index for all interest-bearing government bonds with a maturity of at least one year and calculates the monthly government bond performance index (GBPI) for the period from June 1924 to August 1945 in accordance with the standard methods of modern bond indices. There are two types of GBPI: the PI and the TRI that also reflects income returns. Domestic government bonds for the period for index calculation can be classified into 20 types, and the total number of issues is 268.\footnote{Indexes are calculated for 5\% Loan, 5\% Loan (Special), 5\% Loan (Mark Ko), 4\% Loan (1st issue), 4\% Loan (2nd issue), 4\% Loan, 3.5\% Loan, 5\% Treasury Bonds, Railway Bonds, Extraordinary Treasury Bonds (Issued at Discount), 4.5\% Treasury Bonds, 4\% Treasury Bonds, 3.5\% Treasury Bonds, China Incident (Second Sino-Japanese War) Treasury Bonds, China Incident Treasury Bonds (Special), The Great East Asia War (Pacific War) Treasury Bonds, The Great East Asia War (Pacific War) Treasury Bonds (Special), and Grants Treasury Bonds Delivered.} Hirayama refers to the following resources for calculating the GBPI: the JGB price quotes specified in the monthly average prices in the Tokyo market stated in the {\it{Debt Management Reports}} (Ministry of Finance, Japan) from June 1924 to March 1942, and the spot transaction and monthly average price stated in the {\it{Monthly Statistical Reports}} (Tokyo/Japan Stock Exchange) from April 1942 to November 1944. Because of tighter price controls after December 1944, the bond prices were fixed, and November 1944 values were adopted.

\subsection{Dataset on the Exchange Rate}
Data before 1942 are official prices listed in {\it{Bulletin of Financial Information}} (Ministry of Finance, Japan) for each fiscal year, and those after 1942 are official prices listed in \citet{boj1947hsw}.

\bigskip

\bigskip

\bigskip

\setcounter{table}{0}
\renewcommand{\thetable}{\arabic{table}}

\clearpage

\begin{table}[tbp]
\caption{Descriptive Statistics and Unit Root Tests}
\label{semi_strong_table1}
\begin{center}
\begin{tabular}{llccccc}\hline\hline
 &  & & EQPI & GBPI & Exchange & \\\hline
 & Mean & & 0.0051  & 0.0046 & 0.0021  & \\
 & SD & & 0.0462  &  0.0058 & 0.0299  & \\
 & Min & & $-0.1580$  & $-0.0362$  & $-0.1133$  & \\
 & Max & & 0.1991  &  0.0293  & 0.3229  & \\\hline
 & ADF-GLS & & $-11.0501$   & $-9.7936$  & $-7.3123$  & \\
 & Lags & & 0  & 0  & 0  & \\
 & $\hat{\phi}$ & & 0.3306  & 0.4340  & 0.3215  & \\\hline
 & $\mathcal{N}$ & & 245 & 245 & 245 & \\\hline\hline
\end{tabular}
\vspace*{5pt}
{
\begin{minipage}{350pt}\footnotesize
{\underline{Notes:}}
\begin{itemize}
\item[(1)] ``EQPI'', ``GBPI'' and ``Exchange'' denote the returns on the equity performance index, the government bond performance index and the exchange rate, respectively.
\item[(2)] ``ADF-GLS'' denotes the ADF-GLS test statistics, ``Lags'' denotes the lag order selected by the MBIC, and ``$\hat\phi$'' denotes the coefficients vector in the GLS detrended series (see Equation (6) in \citet{ng2001lls}).
\item[(3)] In the ADF-GLS test, a model with a time trend and a constant is assumed. The critical value at the 5\% significance level for the ADF-GLS test is ``$-2.91$''.
\item[(4)] ``$\mathcal{N}$'' denotes the number of observations.
\item[(5)] R version 4.1.0 was used to compute the statistics.
\end{itemize}
\end{minipage}}%
\end{center}
\end{table}

\pagebreak

\begin{table}[tbp]
\caption{Preliminary Estimation and Parameter Constancy Test}
\label{semi_strong_table2}
\begin{center}
\begin{tabular}{ccccccc}\hline\hline
 &  & EQPI & GBPI & Exchange & \\\hline
 & \multirow{2}*{$Constant$} & $0.0031$  & $0.0024$  & $0.0004$  & \\
 & & $[0.0042]$ & $[0.0004]$  & $[0.0039]$  & \\
 & \multirow{2}*{$R_{S,t-1}$} & $0.2541$  & $-0.0123$  & $0.0073$  & \\
 & & $[0.0572]$  & $[0.0071]$  & $[0.1054]$  & \\
 & \multirow{2}*{$R_{G,t-1}$} & $0.0991$  & $0.4851$  & $0.2664$  & \\
 & & $[0.6138]$  & $[0.0504]$  & $[0.8110]$  & \\
 & \multirow{2}*{$R_{E,t-1}$} & $0.0623$  & $-0.0045$  & $0.2464$  & \\
 & & $[0.0958]$  & $[0.0117]$  & $[0.1126]$  & \\\hline
 & $\bar{R}^2$ & $0.0625$  & $0.2000$  & $0.0575$  & \\
 & $L_C$ & \multicolumn{3}{c}{$34.2436$}  & \\\hline\hline
\end{tabular}
\vspace*{5pt}
{
\begin{minipage}{350pt}\footnotesize
{\underline{Notes:}}
 \begin{itemize}
  \item[(1)] ``$R_{t-p}$,'' ``$\bar{R}^2$,'' and ``$L_C$'' denote the VAR($p$) estimate, the adjusted $R^2$, and the \citetapos{hansen1992a} joint $L$ statistic with variance, respectively.
  \item[(2)] \citetapos{newey1987sps} robust standard errors are between brackets.
  \item[(3)] R version 4.1.0 was used to compute the estimates.
 \end{itemize}
\end{minipage}}%
\end{center}
\end{table} 

\clearpage

\begin{table}[tbp]
\caption{Major Historical Events in Prewar Japan}
\label{semi_strong_table3}
 \begin{center}\footnotesize
  \begin{tabular}{clcp{10cm}c}\hline\hline
   & Periods & & Major historical events & \\\hline
   & January 1926 & & TSE's new share price skyrocketed with the announcement of additional payment (from 12.50 yen to 25.00 yen). & \\
   & April 1927 & & The nationwide financial panic sparked when debates in the Diet revealed financial difficulties between the Bank of Taiwan and Suzuki \& Co. (The Showa Financial Crisis); The amended Banking Law promulgated on March 30. & \\
   & July 1929 & & The Hamaguchi Cabinet announced the time for the lifting of an embargo on the export of gold. & \\
   & October 1929 & & The Great Crash occurred. & \\\hline
   & January 1930 & & The Japanese government lifted an embargo on the export of gold. & \\
   & November 1930 & & Seiho-Shouken (Stock Price Keeping Organization cooperated by Japan's Life Insurance Companies) established. & \\
   & September 1931 & & The U.K. abolished the gold standard on 21 September; With the rise of dollar buying speculation in the foreign exchange market, the value of government bonds and stocks declined. & \\
   & November 1931 & & Expectations for a re-ban on gold exports rose, and stock prices started to reverse and rise. & \\
   & December 1931 & & The Inukai Cabinet abandoned the gold standard on 13 December; The TSE's new share price skyrocketed with the announcement of additional payment (from 25.00 yen to 37.50 yen). In contrast to Japan, stock prices in the U.S., U.K., and France were declining. & \\
   & January 1932 & & The official exchange rate of the Japanese yen against the U.S. dollar rapidly decreased. & \\ 
   & July 1932 & & The Capital Flight Prevention Law and The Act on the Calculation of Government Bond Prices promulgated and enforced on 1 July & \\
   & November 1932 & & The Bank of Japan started underwriting government bonds. & \\
   & May 1933 & & The Foreign Exchange Control Law enforced. & \\ 
   & February 1936 & & February 26 incident occurred; Korekiyo Takahashi's economic policy ended. & \\
   & July -- August 1937 & & The Second Sino-Japanese War occurred. & \\\hline
   & July 1941 & & President Franklin Roosevelt froze all Japanese assets in the U.S. & \\
   & December 1941 & & The Pacific War occurred; The Ministry of Commerce and Industry transferred the administrative work of the exchange to the Ministry of Finance. & \\
   & June 1942 & & The Battle of Midway occurred. & \\
   & December 1942 & & The Japanese government decided to delist the TSE shares. & \\
   & February 1943 & & The Operation {\it{Ke}} completed (withdrawal of Japanese forces from Guadalcanal). & \\
   & July 1944 & & The Operation Forager (the Battle of Saipan) completed. & \\
   & July/August 1944 & & The Wartime Finance Bank started the stock price-keeping operation. & \\
   & March 1945 & & The Great Tokyo Air Raids occurred; The Wartime Finance Bank started the stock price-keeping operation (unlimited stock purchase). In July, the Japan stock exchange started the stock price-keeping operation. & \\\hline\hline
 \end{tabular}
 \vspace*{5pt}
 {
 \begin{minipage}{420pt}\footnotesize
  \underline{Note}: This table is constructed following \citet{boj1993cfm} and \citet{fukami2012his}.
 \end{minipage}}%
 \end{center}
 \end{table}

\clearpage

\clearpage

\clearpage

\begin{thebibliography}{72}
\newcommand{\enquote}[1]{``#1''}
\expandafter\ifx\csname natexlab\endcsname\relax\def\natexlab#1{#1}\fi

\bibitem[{Ball and Brown(1968)}]{ball1968eea}
Ball, R. and Brown, P. (1968), \enquote{An Empirical Evaluation of Accounting
  Income Numbers,} \textit{Journal of Accounting Research}, 6, 159--178.

\bibitem[{{Bank of Japan}(1947)}]{boj1947hsw}
{Bank of Japan} (1947), \enquote{Senji Chu Tokei Yoran {\rm{[Handbook of
  Statistics During the War]}},} Tokyo, Japan.

\bibitem[{Bassino and Lagoarde-Segot(2015)}]{bassino2015iet}
Bassino, J. and Lagoarde-Segot, T. (2015), \enquote{Informational Efficiency in
  the Tokyo Stock Exchange, 1931--40,} \textit{Economic History Review}, 68,
  1226--1249.

\bibitem[{Becker et~al.(1996)Becker, Finnerty, and Kopecky}]{becker1996mne}
Becker, K.~G., Finnerty, J.~E., and Kopecky, K.~J. (1996),
  \enquote{Macroeconomic News and the Efficiency of International Bond Futures
  Markets,} \textit{Journal of Futures Markets}, 16, 131--145.

\bibitem[{Bernanke(2004)}]{bernanke2004wpl}
Bernanke, B.~B. (2004), \enquote{What Policymakers can Learn from Asset
  Prices,} Tech. rep., Remarks before The InvestmentAnalysts Society of
  Chicago.

\bibitem[{Brown and Warner(1980)}]{brown1980msp}
Brown, S.~J. and Warner, J.~B. (1980), \enquote{Measuring Security Price
  Performance,} \textit{Journal of Financial Economics}, 8, 205--258.

\bibitem[{Campbell(1987)}]{campbell1987srt}
Campbell, J.~Y. (1987), \enquote{Stock Returns and the Term Structure,}
  \textit{Journal of Financial Economics}, 18, 373--399.

\bibitem[{Campbell and Shiller(1988{\natexlab{a}})}]{campbell1988dpr}
Campbell, J.~Y. and Shiller, R.~J. (1988{\natexlab{a}}), \enquote{The
  Dividend-Price Ratio and Expectations of Future Dividends and Discount
  Factors,} \textit{Review of Financial Studies}, 1, 195--228.

\bibitem[{Campbell and Shiller(1988{\natexlab{b}})}]{campbell1988spe}
--- (1988{\natexlab{b}}), \enquote{Stock Prices, Earnings, and Expected
  Dividends,} \textit{Journal of Finance,}, 43, 661--676.

\bibitem[{Choudhry(2010)}]{choudhry2010ww2}
Choudhry, T. (2010), \enquote{World War I\hspace{-.1em}I Events and the Dow
  Jones Industrial Index,} \textit{Journal of Banking and Finance}, 34,
  1022--1031.

\bibitem[{Chuli{\'a} et~al.(2010)Chuli{\'a}, Martens, and van
  Dijk}]{chulia2010aef}
Chuli{\'a}, H., Martens, M., and van Dijk, D. (2010), \enquote{Asymmetric
  Effects of Federal Funds Target Rate Changes on S\&P100 Stock Returns,
  Volatilities and Correlations,} \textit{Journal of Banking \& Finance}, 34,
  834--839.

\bibitem[{Darrat(1988)}]{darrat1988fps}
Darrat, A.~F. (1988), \enquote{On Fiscal Policy and the Stock Market,}
  \textit{Journal of Money, Credit and Banking}, 20, 353--363.

\bibitem[{Davidson and Froyen(1982)}]{davidson1982mps}
Davidson, L.~S. and Froyen, R.~T. (1982), \enquote{Monetary Policy and Stock
  Returns:Are Stock Markets Efficient?} \textit{Federal Reserve Bank of St.
  Louis Review}, 64, 3--12.

\bibitem[{Ehrmann and Fratzscher(2004)}]{ehrmann2004tsm}
Ehrmann, M. and Fratzscher, M. (2004), \enquote{Taking Stock: Monetary Policy
  Transmission to Equity Markets,} \textit{Journal of Money, Credit and
  Banking}, 36, 719--737.

\bibitem[{Elliott et~al.(1996)Elliott, Rothenberg, and Stock}]{elliott1996eta}
Elliott, G., Rothenberg, T.~J., and Stock, J.~H. (1996), \enquote{Efficient
  Tests for an Autoregressive Unit Root,} \textit{Econometrica}, 64, 813--836.

\bibitem[{Fama(1970)}]{fama1970ecm}
Fama, E.~F. (1970), \enquote{Efficient Capital Markets: A Review of Theory and
  Empirical Work,} \textit{Journal of Finance}, 25, 383--417.

\bibitem[{Fama(1981)}]{fama1981srr}
--- (1981), \enquote{Stock Returns, Real Activity, Inflation, and Money,}
  \textit{American Economic Review}, 71, 545--565.

\bibitem[{Fama(1991)}]{fama1991ecm}
--- (1991), \enquote{Efficient Capital Markets: {I\hspace{-.1em}I},}
  \textit{Journal of Finance}, 46, 1575--1617.

\bibitem[{Fama et~al.(1969)Fama, Fisher, Jensen, and Roll}]{fama1969asp}
Fama, E.~F., Fisher, L., Jensen, M.~C., and Roll, R. (1969), \enquote{The
  Adjustment of Stock Prices to New Information,} \textit{International
  Economic Review}, 10, 1--21.

\bibitem[{Fama and French(1988)}]{fama1988dye}
Fama, E.~F. and French, K.~R. (1988), \enquote{Dividend Yields and Expected
  Stock Returns,} \textit{Journal of Financial Economics}, 22, 3--25.

\bibitem[{Fama and French(1989)}]{fama1989bce}
--- (1989), \enquote{Business Conditions and Expected Returns on Stocks and
  Bonds,} \textit{Journal of Financial Economics}, 25, 23--49.

\bibitem[{Fujino and Teranishi(2000)}]{fujino2000qaj}
Fujino, S. and Teranishi, J. (2000), \textit{A Quantitative Analysis of the
  Japanese Financial System}, Toyo Keizai Shimpo Sha.

\bibitem[{Fukai(1941)}]{fukai1941r70}
Fukai, E. (1941), \textit{Kaiko 70-nen \rm{[Reflections of 70 Years]}}, Iwanami
  Shoten, Tokyo.

\bibitem[{Fukami(2012)}]{fukami2012his}
Fukami, Y. (2012), \enquote{History of Institute for Stock Price- Keeping in
  Prewar Showa Period Focusing on the Seiho-Shouken (in Japanese),}
  \textit{Journal of Financial and Securities Markets {{\rm (Japan Securities
  Research Institute)}}}, 78, 1--18.

\bibitem[{Hancock(1989)}]{hancock1989fpm}
Hancock, D.~G. (1989), \enquote{Fiscal Policy, Monetary Policy and the
  Efficiency of the Stock Market,} \textit{Economics Letters}, 31, 65--69.

\bibitem[{Hansen(1992)}]{hansen1992a}
Hansen, B.~E. (1992), \enquote{Testing for Parameter Instability in Linear
  Models,} \textit{Journal of Policy Modeling}, 14, 517--533.

\bibitem[{Harada and Okimoto(2019)}]{harada2019boj}
Harada, K. and Okimoto, T. (2019), \enquote{The BOJ's ETF Purchases and Its
  Effects on Nikkei 225 Stocks,} RIETI Discussion Paper Series 19-E-014.

\bibitem[{Hirayama(2017{\natexlab{a}})}]{hirayama2017jsm}
Hirayama, K. (2017{\natexlab{a}}), \enquote{The Japanese Stock Market
  Performance Index in the Early Showa Era (in Japanese),} \textit{Annals of
  Society for the Economic Studies of Securities {\rm{(Japan Securities
  Research Institute)}}}, 51, 1--12.

\bibitem[{Hirayama(2017{\natexlab{b}})}]{hirayama2017rje}
--- (2017{\natexlab{b}}), \enquote{Reappraisal of Japanese Equity Market Return
  in the Early Showa Era (in Japanese),} \textit{Journal of Economic Science
  {\rm{(The Economics Society of Saitama University)}}}, 14, 41--53.

\bibitem[{Hirayama(2018{\natexlab{a}})}]{hirayama2018erj}
--- (2018{\natexlab{a}}), \enquote{The Japanese Equity Performance Index in the
  Early Showa Era (in Japanese),} \textit{Journal of Financial and Securities
  Markets {\rm{(Japan Securities Research Institute)}}}, 101, 71--91.

\bibitem[{Hirayama(2018{\natexlab{b}})}]{hirayama2018jgb}
--- (2018{\natexlab{b}}), \enquote{The Japanese Government Bond Performance
  Index in the Early Showa Era (in Japanese),} \textit{Review of Monetary and
  Financial Studies {\rm{(Japan Society of Monetary Economics)}}}, 40, 54--65.

\bibitem[{Hirayama(2019)}]{hirayama2019lpp}
--- (2019), \enquote{A Linkage between Prewar and Postwar Price in the Japanese
  Stock Market,} Institute for Stock Price Index Workshop at Meiji University
  (August 19, 2019).

\bibitem[{Hirayama(2020)}]{hirayama2020jep}
--- (2020), \enquote{The Japanese Equity Performance from 1944 to 1945 (in
  Japanese),} \textit{Journal of Financial and Securities Markets {{\rm (Japan
  Securities Research Institute)}}}, 109, 63--85.

\bibitem[{Hirayama(2021)}]{hirayama2021jgb}
--- (2021), \enquote{The Japanese Government Bond Performance from 1944 to 1945
  (in Japanese),} \textit{Journal of Financial and Securities Markets
  {\rm{(Japan Securities Research Institute)}}}, 113, 1--27.

\bibitem[{Hirayama and Noda(2019)}]{hirayama2019mtv}
Hirayama, K. and Noda, A. (2019), \enquote{Measuring the Time-Varying Market
  Efficiencyin the Prewar Japanese Stock Market,} [arXiv:1911.04059], Available
  at https://arxiv.org/pdf/1911.04059.pdf.

\bibitem[{Hoshi and Kashyap(2001)}]{hoshi2001cfg}
Hoshi, T. and Kashyap, A. (2001), \textit{Corporate Finance and Governance in
  Japan: The Road to the Future}, MIT Press.

\bibitem[{{Institute for Monetary and Economic Studies, Bank of
  Japan}(1993)}]{boj1993cfm}
{Institute for Monetary and Economic Studies, Bank of Japan} (1993),
  \enquote{Chronology of Financial Matters in Japan {\rm{(in Japanese)}},}
  {Institute for Monetary and Economic Studies, Bank of Japan}, revised
  Edition.

\bibitem[{Ishii(1997)}]{ishii1997irj}
Ishii, K. (1997), \textit{Nihon no Sangyo Kakumei {\rm{[Industrial Revolution
  in Japan]}}}, Asahi Sensho.

\bibitem[{Ishii(1999)}]{ishii1999tfh}
--- (1999), \textit{Nihon no Sangyo Kakumei (Industrial revolution in
  Japan).Kindai Nohon Kinyushi Josetsu {\rm{[Towards the Financial History of
  Modern Japan]}}}, The University of Tokyo Press.

\bibitem[{Ito(1989)}]{ito1989jof}
Ito, M. (1989), \textit{Nihon no Taigai Kin'yu to Kin'yu Seisaku {\rm{(Japan's
  Overseas Finances and Monetary Policies: 1914-1936)}}}, Nihon Keizai Hyoron
  Sha, Tokyo.

\bibitem[{Ito et~al.(2016)Ito, Maeda, and Noda}]{ito2016meg}
Ito, M., Maeda, K., and Noda, A. (2016), \enquote{Market Efficiency and
  Government Interventions in Prewar Japanese Rice Futures Markets,}
  \textit{Financial History Review}, 23, 325--346.

\bibitem[{Ito et~al.(2018)Ito, Maeda, and Noda}]{ito2018fpe}
--- (2018), \enquote{The Futures Premium and Rice Market Efficiency in Prewar
  Japan,} \textit{Economic History Review}, 71, 909--937.

\bibitem[{Ito et~al.(2014)Ito, Noda, and Wada}]{ito2014ism}
Ito, M., Noda, A., and Wada, T. (2014), \enquote{International Stock Market
  Efficiency: A Non-Bayesian Time-Varying Model Approach,} \textit{Applied
  Economics}, 46, 2744--2754.

\bibitem[{Ito et~al.(2017)Ito, Noda, and Wada}]{ito2017aae}
--- (2017), \enquote{An Alternative Estimation Method of a Time-Varying
  Parameter Model,} [arXiv:1707.06837], Available at
  https://arxiv.org/pdf/1707.06837.pdf.

\bibitem[{Ito(1995)}]{ito1995hsj}
Ito, O. (1995), \textit{Nihon-gata Kinyu no Rekishi-teki Kozo {\rm{[The
  Historical Structure of Japan's Financial System]}}}, University of Tokyo
  Press.

\bibitem[{Kasuya(2006)}]{kasuya2006sir}
Kasuya, M. (2006), \enquote{Securities Investments of Regional Banks during
  Interwar Period Japan,} \textit{Monetary and Economic Studies}, 25, 59--104.

\bibitem[{Kataoka et~al.(2004)Kataoka, Maru, and Teranishi}]{kataoka2004b}
Kataoka, Y., Maru, J., and Teranishi, J. (2004), \enquote{An Analysis of Stock
  Market Efficiency in the Late Meiji Era, Part.2 {\rm{(in Japanese)}},}
  \textit{Journal of Financial and Securities Markets {\rm{(Japan Securities
  Research Institute)}}}, 48, 69--81.

\bibitem[{Kato(1965{\natexlab{a}})}]{kato1965bcw1}
Kato, T. (1965{\natexlab{a}}), \enquote{Banking Conditions under the War
  Economy : Through an Analysis of Big Banks (I),} \textit{Journal of Social
  Science}, 17, 1--42.

\bibitem[{Kato(1965{\natexlab{b}})}]{kato1965bcw2}
--- (1965{\natexlab{b}}), \enquote{Banking Conditions under the War Economy :
  Through an Analysis of Big Banks (I),} \textit{Journal of Social Science},
  17, 94--119.

\bibitem[{Kobayashi(2012)}]{kobayashi2012hjs}
Kobayashi, K. (2012), \textit{Nihon Shoken Shiron {\rm{(History of Japanese
  Securities)}}}, Nihon Keizai Hyoron Sha, Tokyo.

\bibitem[{Laopodis(2009)}]{laopodis2009fps}
Laopodis, N.~T. (2009), \enquote{Fiscal Policy and Stock Market Efficiency:
  Evidence for the United States,} \textit{Quarterly Review of Economics and
  Finance}, 49, 633--650.

\bibitem[{Lim and Brooks(2011)}]{lim2011esm}
Lim, K.~P. and Brooks, R. (2011), \enquote{The Evolution of Stock Market
  Efficiency Over Time: A Survey of the Empirical Literature,} \textit{Journal
  of Economic Surveys}, 25, 69--108.

\bibitem[{Lo(2004)}]{lo2004amh}
Lo, A.~W. (2004), \enquote{The Adaptive Markets Hypothesis: Market Efficiency
  from an Evolutionary Perspective,} \textit{Journal of Portfolio Management},
  30, 15--29.

\bibitem[{MacKinlay(1997)}]{mackinlay1997ese}
MacKinlay, A.~C. (1997), \enquote{Event Studies in Economics and Finance,}
  \textit{Journal of Economic Literature}, 35, 13--39.

\bibitem[{Malkiel(2003)}]{malkiel2003emh}
Malkiel, B.~G. (2003), \enquote{The Efficient Market Hypothesis and its
  Critics,} \textit{Journal of Economic Perspectives}, 17, 59--82.

\bibitem[{McWilliams and Siegel(1997)}]{mcwilliams1997esm}
McWilliams, A. and Siegel, D. (1997), \enquote{Event Studies in Management
  Research: Theoretical and Empirical Issues,} \textit{Academy of Management
  Journal}, 40, 626--657.

\bibitem[{Newey and West(1987)}]{newey1987sps}
Newey, W.~K. and West, K.~D. (1987), \enquote{A Simple, Positive Semi-Definite,
  Heteroskedasticity and Autocorrelation Consistent Covariance Matrix,}
  \textit{Econometrica}, 55, 703--708.

\bibitem[{Ng and Perron(2001)}]{ng2001lls}
Ng, S. and Perron, P. (2001), \enquote{Lag Length Selection and the
  Construction of Unit Root Tests with Good Size and Power,}
  \textit{Econometrica}, 69, 1519--1554.

\bibitem[{Okazaki and Okuno-Fujiwara(1993)}]{okazaki1993hoc}
Okazaki, T. and Okuno-Fujiwara, M. (1993), \textit{Gendai Nihon Keizai
  Shisutemu no Genryu \rm [Historical Origins of the Contemporary Japanese
  Economic System]}, Nihon Keizai Shinbunsha.

\bibitem[{Pearce and Roley(1985)}]{pearce1985spe}
Pearce, D. and Roley, V. (1985), \enquote{Stock Prices and Economic News,}
  \textit{Journal of Business}, 58, 49--67.

\bibitem[{Perron(1989)}]{perron1989tgc}
Perron, P. (1989), \enquote{The Great Crash, the Oil Price Shock, and the Unit
  Root Hypothesis,} \textit{Econometrica}, 57, 1361--1401.

\bibitem[{Schwarz(1978)}]{schwarz1978edm}
Schwarz, G. (1978), \enquote{Estimating the Dimension of a Model,}
  \textit{Annals of Statistics}, 6, 461--464.

\bibitem[{Shibamoto and Shizume(2014)}]{shibamoto2014era}
Shibamoto, M. and Shizume, M. (2014), \enquote{Exchange Rate Adjustment,
  Monetary Policy and Fiscal Stimulus in Japan's Escape from the Great
  Depression,} \textit{Explorations in Economic History}, 53, 1--18.

\bibitem[{Shibata(2011)}]{shibata2011mcw}
Shibata, Y. (2011), \textit{Senji Nihon no Kinyu Tosei {\rm{[Monetary Controls
  in Wartime Japan]}}}, Nihon Keizai Hyoron Sha, Tokyo.

\bibitem[{Shiller(1981)}]{shiller1981dsp}
Shiller, R.~J. (1981), \enquote{{Do Stock Prices React Too Much to Be Justified
  by Subsequent Changes in Dividends?}} \textit{American Economic Review}, 71,
  421--436.

\bibitem[{Suzuki(2012)}]{suzuki2012pwt}
Suzuki, S. (2012), \enquote{Pacific War and Tokyo Stock Exchange Daily
  Data:1941-1943 {\rm{(in Japanese)}},} \textit{Bulletin of Economic Studies
  {\rm{(Meisei University)}}}, 44, 39--51.

\bibitem[{Suzuki and Yuki(2019)}]{suzuki2019gke}
Suzuki, S. and Yuki, T. (2019), \enquote{Great Kanto Earthquake and the
  Japanese Stock Market {\rm{(in Japanese)}},} Mimeo.

\bibitem[{Teranishi(2003)}]{teranishi2003esj}
Teranishi, J. (2003), \textit{Nihon no Keizai Shisutemu {\rm{[The Economic
  System in Japan]}}}, Iwanami Shoten.

\bibitem[{Teranishi(2011)}]{teranishi2011fsp}
--- (2011), \textit{Senzenki Nihon no Kinyu Shisutemu {\rm{[The Financial
  System in Prewar Japan]}}}, Iwanami Shoten.

\bibitem[{Ueda(2012)}]{ueda2012enm}
Ueda, K. (2012), \enquote{The Effectiveness of Non-traditional Monetary Policy
  Measures: The Case of the Bank of Japan,} \textit{Japanese Economic Review},
  63, 1--22.

\bibitem[{Ueda(2013)}]{ueda2013ra}
--- (2013), \enquote{Response of Asset Prices to Monetary Policy under
  Abenomics,} \textit{Asian Economic Policy Review}, 8, 252--269.

\bibitem[{Utsunomiya(2013)}]{utsunomiya2013jfi}
Utsunomiya, K. (2013), \enquote{Japan's Financial Intermediation in the 1940s:
  Estimation of Flow of Funds Accounts from 1941 to 1948 (in Japanese),}
  \textit{Review of Monetary and Financial Studies {\rm{(Japan Society of
  Monetary Economics)}}}, 35, 52--73.

\end{thebibliography}
\end{document}